\documentclass[preprint2]{aastex}

\shorttitle{SN 2002cx-like and SN Ia-CSM objects}
\shortauthors{Meng \& Podsiadlowski}

\begin{document}


\title{Do SN 2002cx-like and SN Ia-CSM objects share the same origin?}


\author{X. Meng$^{\rm 1,2,3}$ \& Ph. Podsiadlowski$^{\rm 4}$}
\affil{$^{\rm 1}$Yunnan Observatories, Chinese Academy of Sciences, 650216 Kunming, PR China\\
$^{\rm 2}$ Key Laboratory for the Structure and Evolution of
Celestial Objects, Chinese Academy of Sciences, 650216 Kunming, PR
China\\
$^{\rm 3}$Center for Astronomical Mega-Science, Chinese Academy of
Sciences, 20A Datun Road, Chaoyang District, Beijing, 100012, PR China \\
$^{\rm 4}$Department of Astronomy, Oxford University, Oxford OX1
3RH, UK} \email{xiangcunmeng@ynao.ac.cn}





\begin{abstract}
SN 2002cx-like and SN Ia-CSM objects show similar early spectra
and both belong to a young stellar population, suggesting that
they could share the same progenitor origin.  Adopting the
framework of the common-envelope-wind (CEW) model developed in
\citet{MENGXC17}, we here propose that both subclasses of SNe Ia
are caused by the explosion of hybrid carbon-oxygen-neon white
dwarfs (CONe WDs) in single-degenerate systems, where SNe Ia-CSM
explode in systems with a massive common envelope (CE) of
$\sim1~M_{\odot}$, while SN 2002cx-like events correspond to those
events where most of the CE has been lost in a wind. Using
binary-population-synthesis (BPS) calculations, we estimate a
number ratio of SNe Ia-CSM to SN 2002cx-like objects between
$\frac{1}{3}$ and $\frac{2}{3}$, consistent with observational
constraints, and an overall contribution from hybrid CONe WDs to
the total SN Ia population that also matches the observed number
from these peculiar objects. Our model predicts a statistical
sequence of CSM density from SN Ia-CSM to SN 2002cx-like events
and normal SNe Ia, consistent with existing radio constraints. We
also find a new subclass of hybrid SNe which share the properties
of Type II and Type Ia SNe, consistent with some observed SNe,
which do not have a surviving companion. In some cases these could
even produce SNe Ia from apparently single WDs.
\end{abstract}





\section{INTRODUCTION}
\label{sect:1} Although Type Ia supernovae (SNe Ia) are known to
be astrophysical events of major importance, e.g.\ as standard
candles to measure cosmological parameters (\citealt{RIE98};
\citealt{PER99}), the exact nature of their progenitors has
remained unclear (\citealt{HN00}; \citealt{LEI00}). There is a
consensus that a SN Ia results from the thermonuclear explosion of
a carbon/oxygen white dwarf (CO WD) in a binary system
(\citealt{HF60}), but there is still a decade-long debate
concerning the nature of the companion. Two basic scenarios for
the progenitors of SNe Ia have been discussed for the last four
decades. One is the single-degenerate (SD) model, where the CO WD
accretes material from a non-degenerate companion star
(\citealt{WI73}; \citealt{NTY84}); the other is the
double-degenerate (DD) model, involving the merger of two CO WDs
(\citealt{IT84}; \citealt{WEB84}). At present, support and
counter-arguments exist for the two basic scenarios
(\citealt{WANGB12}; \citealt{MAOZ14}) both on the observational
and the theoretical side.

The detection of circumstellar material (CSM) in the spectrum of
SNe Ia is usually taken as strong evidence in favour of the SD
model for these objects (\citealt{DILDAY12}; \citealt{MAGUIRE13}).
Especially, a new subclass of SN Ia-like objects shows the
spectroscopic signatures of both SNe Ia and IIn in the form of
broad Fe, Ca, S and Si absorption lines with strong narrow
H$\alpha$ emission lines, which is explained by a SN Ia exploding
in a dense CSM. The interaction between the supernova ejecta and
the CSM partly contributes to their high luminosity. The first
candidate for this subclass was SN 2002ic, which is characterized
by the absorption features seen in SN 1991T-like SNe Ia and strong
H$\alpha$ emission lines (\citealt{HAMUY03}). Such SNe Ia are
referred to by various authors as SN Ia/IIn, Ian, IIa, IIan or SNe
Ia-CSM (\citealt{SILVERMAN13}).  Here, we denote such events as
``SNe Ia-CSM'' following \citet{SILVERMAN13}. Although there is
some debate on whether these objects are truly
SNe Ia or in fact core-collapse SNe (\citealt{BENETTI06};
\citealt{INSERRA14}), the discovery of PTF11kx definitively shows
that at least some of the SN Ia-CSM events are connected with SNe
Ia (\citealt{DILDAY12}). In addition, the host galaxies of all SNe
Ia-CSM are late-type spirals, which implies that these objects
originate from a relatively young stellar population
(\citealt{SILVERMAN13}). Within the common-envelope wind (CEW)
model, \citet{MENGXC17} suggested that 2002ic-like SNe are
connected with the explosion of CO WDs in a massive CE, but this
interpretation seems to have difficulties in explaining the event
rate for the whole SN Ia-CSM class.

SN 2002cx has been called the most peculiar known SN Ia
(\citealt{LIWD02}). Similarly to SN Ia-CSM objects, 2002cx-like
SNe also exhibit SN 1991T-like pre-maximum spectra
(\citealt{LIWD02}). Moreover, just as SN Ia-CSM objects,
2002cx-like SNe favour late-type galaxies (\citealt{FOLEY13};
\citealt{LYMAN18}). In addition, although radio observations
provide only an upper limit on the CSM density, an analysis of the
upper limit based on an analytical models for the temporal and
spectral evolution of prompt radio emission from the interaction
with the CSM shows that, compared with normal SNe Ia, SN
2002cx-like events seem to have a relatively dense CSM, but not as
dense as SNe Ia-CSM (e.g. Fig.~7 in \citealt{CHOMIUK16}). Does the
similarity between SN 2002cx-like and SN Ia-CSM objects imply that
they share the same progenitor channel? Recently, 2002cx-like SNe
were proposed to be caused by the explosions of hybrid
carbon-oxygen-neon (CONe) WDs in SD systems (\citealt{MENGXC14};
\citealt{WANGB14}); the chemical evolution of dwarf spheroidal
galaxies may even provide some indirect support for this
suggestion, i.e. if there were no contribution of subluminous SNe
Ia from hybrid CONe WDs, the spread of [Mn/Fe] in dwarf spheroidal
galaxies could not be reproduced (\citealt{KOBAYASHI15};
\citealt{CESCUTTI17}). Furthermore, numerical simulations have
demonstrated that the explosions of hybrid CONe WDs could
reproduce their properties, e.g. their low luminosity, low kinetic
energy, and even their light curve and spectrum
(\citealt{KROMER15}; \citealt{BRAVO16}). If SN 2002cx-like and SNe
Ia-CSM objects share the same origin, could the hybrid CONe WD
channel simultaneously explain the origin of both types? In this
paper, we investigate this question and show that they could in
principle.

In section \ref{sect:2}, we briefly describe our method and present
the results of our calculations in section \ref{sect:3}, followed with
a detailed discussion of the results and conclusions in section
\ref{sect:4}.

\begin{figure*}
\centerline{\includegraphics[angle=0,scale=.60]{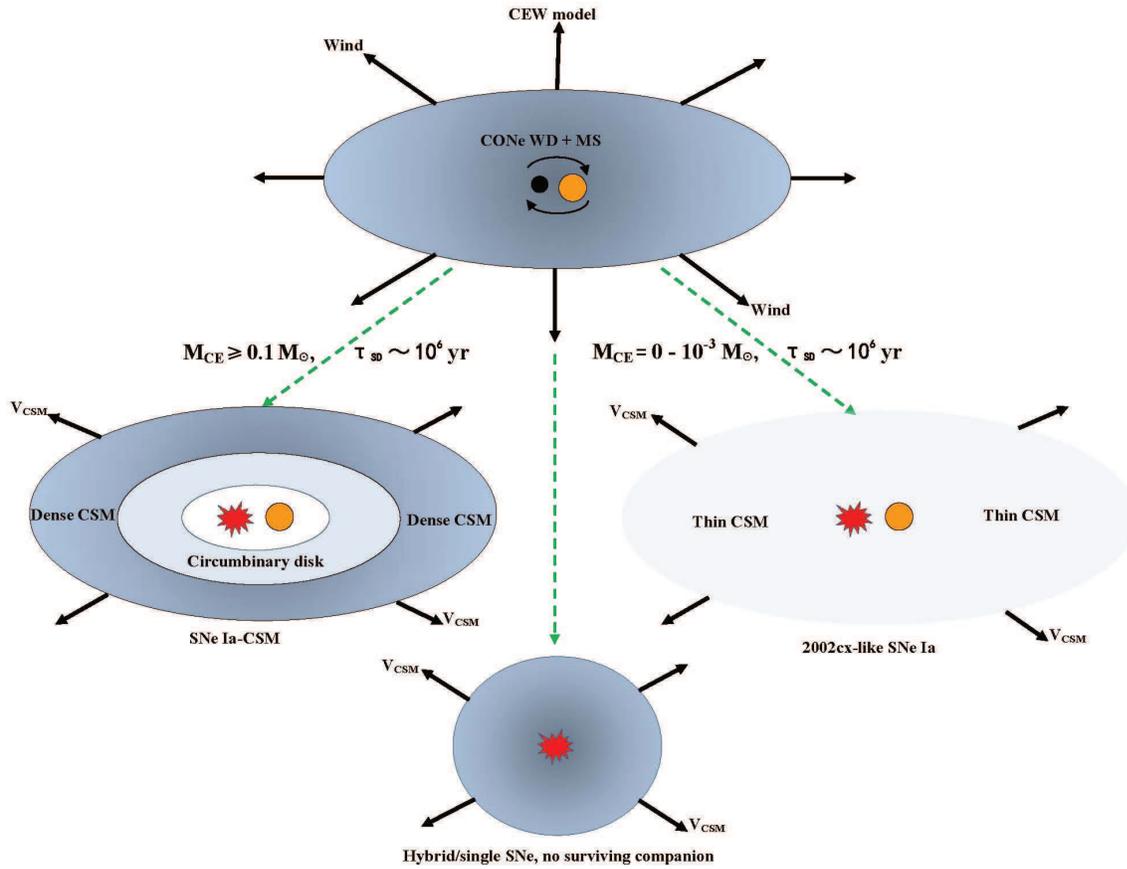}}
\caption{Schematic diagram illustrating the different channels for
forming SN Ia-CSM, SN 2002cx-like and hybrid/single
SNe.}\label{aa}
\end{figure*}

\section{METHOD}
\label{sect:2}
Recently, \citet{MENGXC17} constructed a new version of the SD
model, in which a CE is assumed to form when the mass-transfer
rate between a CO WD and its companion exceeds a critical
accretion rate, rather than the onset of an optically thick wind
(OTW; \citealt{HAC96}). The WD can then gradually increase its mass at
the base of the CE similarly to the degenerate core in a thermally
pulsing asymptotic-giant-branch (TPAGB) star. For the large
nuclear luminosity from stable hydrogen burning, the CE will expand
to giant dimensions and lose mass from the surface of the CE
by a CE wind; this leads to a low CE density and a correspondingly
low frictional luminosity between the binary system and the CE. As
a result, the binary system will avoid a fast spiral-in phase for
a large parameter range and eventually re-emerge from the CE
phase, instead of merging completely. In the CEW model, the SN Ia
may explode in the CE phase, in a phase of stable hydrogen burning
(related to a supersoft X-ray source [SSS] phase) or a phase of
weakly unstable hydrogen burning, where the system would appear as
a recurrent nova [RN]. The CEW model shares many of the merits of
the OTW model while avoiding some of its shortcomings.

In this paper, we calculate the evolution of potential SN Ia
progenitor binaries following the method developed in
\citet{MENGXC17}, except that the WDs are assumed to be hybrid
CONe WDs. Hybrid CONe WDs could be as massive as 1.30 $M_{\odot}$,
which means that WDs do not need to accrete much mass to reach the
Chandrasekhar limit (\citealt{DENISSENKOV13}; \citealt{CHENMC14}).
Here, we only consider the case where the companion is a
main-sequence or a sub-giant star (WD+MS) since the contribution
to the total SN Ia rate from WD binaries with red-giant (RG)
companions is quite uncertain: it is not entirely clear whether a
WD + RG system enters into a dynamically unstable CE or a thermal
timescale CE as required by our CEW model (e.g.\
\citealt{YUNGELSON95}; \citealt{HAC99}; \citealt{HAN04};
\citealt{GE15}; \citealt{LIUDD17}). Throughout this work, `CE'
implies a thermal timescale CE, rather than a dynamically unstable
CE, unless otherwise specified. All WD + MS systems in our study
must have experienced a dynamically unstable CE before they are
formed. We calculated a dense grid of models where we varied the
initial WD masses, secondary masses and orbital periods and
assumed that a SN Ia occurs when $M_{\rm WD}=1.378~ M_{\rm \odot}$
(\citealt{NTY84}). Similarly to what was found in
\citet{MENGXC17}, SNe Ia may explode in CE, SSS or RN phases,
which is the key feature for trying to simultaneously explain the
properties of SN 2002cx-like and SN Ia-CSM objects.

Here, we firstly summarize our proposed scheme on SN 2002cx-like
and SN Ia-CSM objects in a schematic diagram in Fig.\ \ref{aa}. At
the time when $M_{\rm WD}=1.378~M_{\odot}$, the CE mass around the
WD, which provides the mass reservoir to form the CSM, tends to be
bimodal, containing either $\geq0.1~M_{\odot}$ of matter or less
than a few $10^{\rm -3}~M_{\odot}$ (Fig.~\ref{mcedis}). If the
spin-down timescale were $\sim10^{\rm 6}$ yr, the CSM near the
central supernova at the time of the explosion would contain so
little mass that it would be very difficult to directly detect a
signal from the interaction between the supernova ejecta and the
low-mass nearby CSM. However, for systems with massive CEs, the
supernova ejecta will catch up with the dense CSM after several
days (depending on the CE mass and wind velocity, see
Sec.~\ref{sect:4.2} in details). Such an interaction would show
narrow hydrogen emission lines such as seen in the spectra of SNe
Ia-CSM. Otherwise, 2002cx-like SNe are expected if there is no
dense CSM around the supernova. Moreover, some systems are
expected to experience a delayed dynamical instability and merge
soon after $M_{\rm WD}=1.378~M_{\odot}$; this predicts a new
subclass of either hybrid SNe or SNe Ia from single WDs, where the
hybrid objects show features of both SNe II and SNe Ia (such as
suggested for SN 2012ca), but without a surviving companion (see
Sec.~\ref{sect:3.1} in details).

We then performed two binary-population-synthesis (BPS)
simulations adopting this new model grid with the rapid binary
evolution code developed by \citet{HUR00,HUR02}. The Hurley et
al.'s code does not include hybrid CONe WDs. Following
\citet{MENGXC14}, we assumed that, if a WD is less massive than
1.3 $M_{\rm \odot}$ based on the results in \citet{CHENMC14} and
is not a CO WD, it is a hybrid CONe WD. If a binary system in the
simulations evolves to the CONe WD+MS stage and the system is
located in the ($\log P^{\rm i}$, $M_{\rm 2}^{\rm i}$) plane for a
SN Ia at the onset of Roche-lobe overflow (RLOF), we assume that a
SN Ia occurs regardless of how massive the CO core is in the
hybrid WD. We followed the evolution of $10^{\rm 7}$ binaries,
where the primordial binary samples are generated in a Monte-Carlo
way with the following input assumptions: (1) a constant
star-formation rate; (2) the initial mass function (IMF) of
\citet{MS79}; (3) a uniform mass-ratio distribution; (4) a uniform
distribution of separations in $\log a$ for binaries, where $a$ is
the orbital separation; (5) circular orbits for all binaries; (6)
a CE ejection efficiency of $\alpha_{\rm CE}=1.0$ or $\alpha_{\rm
CE}=3.0$, where $\alpha_{\rm CE}$ denotes the fraction of the
released orbital energy used to eject the CE (see
\citealt{MENGXC17} for further details).



\begin{figure*}
\centerline{\includegraphics[angle=270,scale=.65]{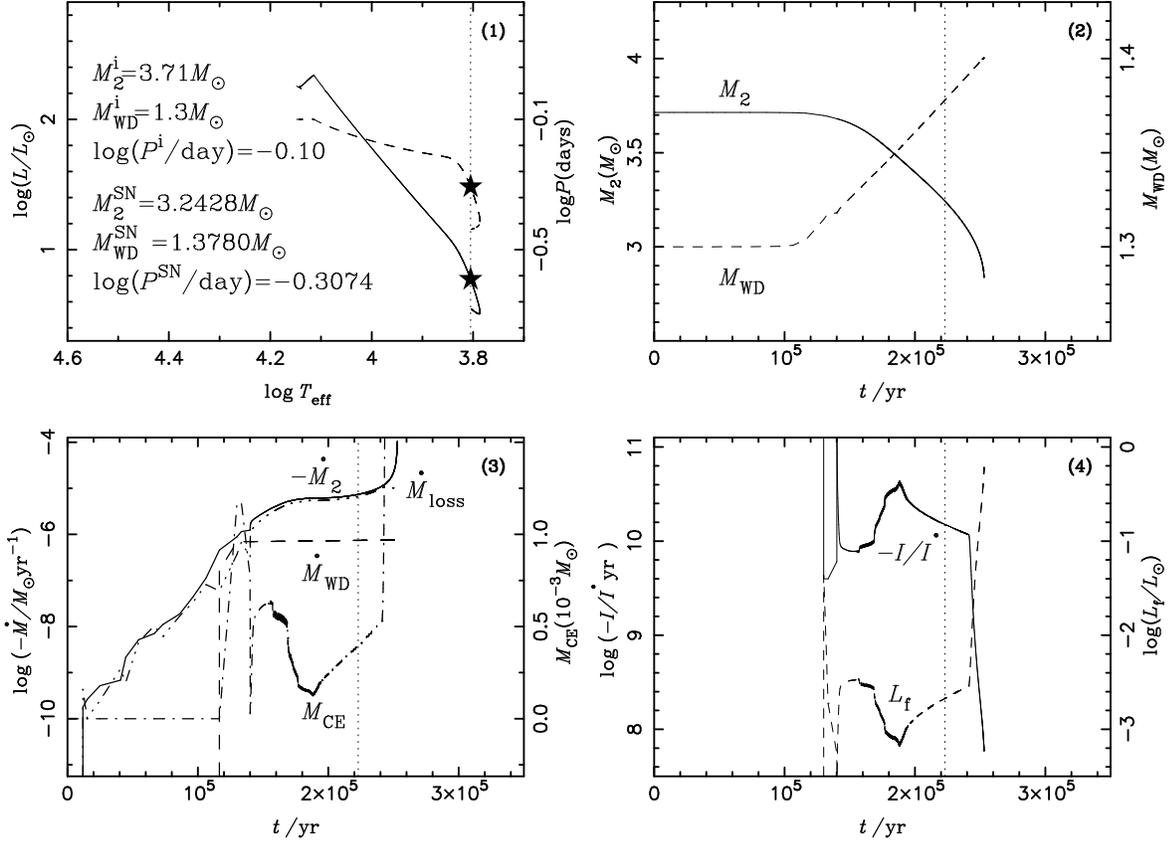}}
\caption{Illustrative binary evolution calculation, where the system
experiences delayed dynamically unstable mass transfer after the
WD has reached $M_{\rm CONe}=1.378~{\rm M_{\rm \odot}}$. The evolution
of various parameters is shown, including the CONe WD mass,
$M_{\rm WD}$, the secondary mass, $M_{\rm 2}$, the mass-transfer
rate, $\dot{M}_{\rm 2}$, the mass-growth rate of the CONe WD,
$\dot{M}_{\rm WD}$, the mass of the CE, $M_{\rm CE}$, the
mass-loss rate from the system, $\dot{M}_{\rm loss}$, the
frictional luminosity, $L_{\rm f}$, and the merger timescale for
the binary system, $I/\dot{I}$, as labelled in each panel. The
evolutionary track of the donor star and the evolution of the
orbital period are shown as solid and dashed curves in panel (1),
respectively. Dotted vertical lines in all panels and asterisks in
panel (1) indicate the position where the WD is expected to
explode as a SN Ia. The initial and the final binary parameters
are given in panel (1).}\label{130137}
\end{figure*}

\section{RESULTS}\label{sect:3}

Generally, the binary evolution sequences for CONe WD + MS systems
are similar to those shown in \citet{MENGXC17}; i.e.\ mass
transfer begins when the companion is a MS star or crosses the
Hertzsprung gap (HG). When $\dot{M}_2$ exceeds a critical
accretion rate, the system enters into the CE phase, and the
hybrid CONe WD increases its mass at the base of the CE. When its
mass reaches 1.378\,$M_{\odot}$, a SN Ia is assumed to occur,
where the system can be in a CE, SSS or RN phase.

\subsection{An example of a delayed dynamical instability}\label{sect:3.1}
However, for the case of $M_{\rm WD}^{\rm i}=1.3~M_{\odot}$, the
evolution of some of the systems differs from the canonical
evolution in an interesting way, an example of which is shown in
Fig.~\ref{130137}. For the system in Fig.~\ref{130137}, the
initial companion is relatively massive and the initial orbital
period is short. Due to the short initial orbital period, the
donor star fills its Roche lobe on the MS, and the system enters
into the CE phase soon thereafter. The WD increases its mass at
the base of the CE, and, after $\sim1.1\times10^{\rm 5}$ yr, the
WD reaches $M_{\rm WD}=1.378~M_{\odot}$, where the CE still
exists.  Since rapidly rotating WDs may explode at a higher mass
than 1.378 $M_{\odot}$ (\citealt{YOON04,YOON05}), we continued our
calculations beyond this mass, assuming the same WD growth pattern
as for $M_{\rm WD}<1.378~M_{\odot}$. We found that after a few
$10^{\rm 4}$ yr, the system will then experience a delayed
dynamical instability, in which the initially stable mass transfer
becomes dynamical unstable later, and then leads to a dynamically
unstable CE. (\citealt{HJELLMING87}).

For a rapidly rotating super-Chandrasekhar WD, it must experience
a spin-down phase before it explodes as a SN Ia
(\citealt{JUSTHAM11}; \citealt{DISTEFANO12}), but the spin-down
timescale is quite uncertain (\citealt{DISTEFANO11};
\citealt{MENGXC13}). For a non-accreting WD, the spin-down
timescale is very likely longer than $10^{\rm 5}$ yr, but less
than a few $10^{\rm 7}$ yr (\citealt{DISTEFANO12};
\citealt{MENGXC13}). So, if the delay time from the point when
$M_{\rm WD}=1.378~M_{\odot}$ to the explosion is longer than the
timescale for the delayed dynamical instability, the system in
Fig.~\ref{130137} would merge before it explodes as a SN Ia,
leaving the WD at the center of the CE and essentially forming a
single AGB star with an overmassive core. If the core is spun down
within its envelope and carbon is ignited at the center of the
hybrid CONe core while there still is a massive envelope around
it, the resulting supernova would have some very unusual
properties (produce a SN I 1/2 in the nomenclature of
\citealt{IBEN83}), with properties between those of a
core-collapse (CC) SN and a SN Ia, an example of which may be
provided by SN 2012ca (\citealt{INSERRA14,INSERRA16}). If the
envelope is ejected before carbon is ignited in the center, it
would lead to the thermonuclear explosion of an apparently {\em
single} CONe WD. In either case, there would be no surviving
companion left after the explosion.

\begin{figure}
\centerline{\includegraphics[angle=270,scale=.35]{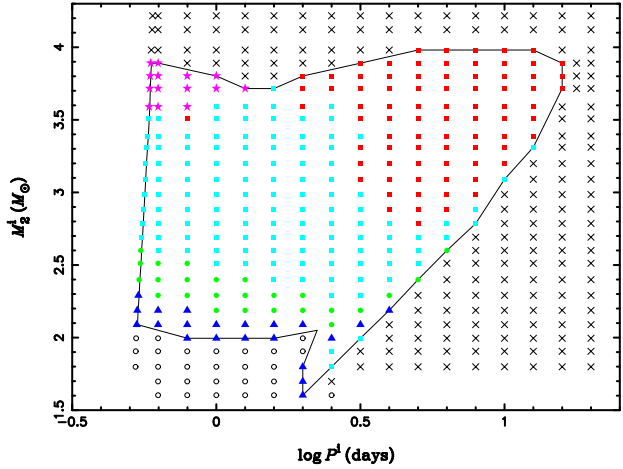}}
\caption{Final outcomes of the binary evolution calculations for
the case of $M_{\rm WD}^{\rm i}=1.3~M_{\odot}$ in the initial
orbital period -- secondary mass ($\log P^{\rm i}, M_{\rm 2}^{\rm
i}$) plane. Filled squares and stars indicate SN Ia explosions
during a CE phase. The red squares indicate that $M_{\rm
CE}\geq0.1~M_{\odot}$, while blue ones denote $M_{\rm
CE}<0.1~M_{\odot}$ when $M_{\rm WD}=1.378~M_{\odot}$.  Filled
stars indicate that the systems will experience a delayed
dynamical instability if the WDs cannot immediately explode as SNe
Ia when $M_{\rm WD}=1.378~M_{\odot}$. Filled circles denote SN Ia
explosions in the SSS phase and filled triangles explosions in the
RN phase. Open circles indicate that the systems experience nova
explosions, preventing the CONe WDs from reaching
1.378\,$M_{\odot}$, while crosses show the systems that are
unstable to dynamical mass transfer. The solid curve gives the
contour of the parameter space for SNe Ia. }\label{grid}
\end{figure}

\begin{figure}
\centerline{\includegraphics[angle=270,scale=.35]{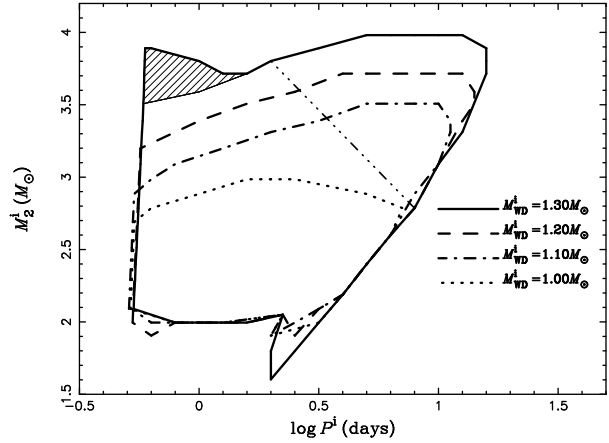}}
\caption{The contours of initial parameters for SNe Ia in the
($\log P^{\rm i},M_{\rm 2}^{\rm i}$) plane for different initial
WD masses. The triple-dot-dashed line roughly shows the boundary
between $M_{\rm CE}\geq0.1~M_{\odot}$ and $M_{\rm
CE}<0.1~M_{\odot}$ when $M_{\rm WD}=1.378~M_{\odot}$. The systems
in the shadowed region will experience a delayed dynamical
instability if the WDs cannot immediately explode as SNe Ia when
$M_{\rm WD}=1.378~M_{\odot}$.}\label{cour130}
\end{figure}

\subsection{Final outcomes of the binary evolution calculations}\label{sect:3.2}
As shown in \citet{MENGXC17}, WDs may explode in a CE, SSS, or RN
phase. When the WD explodes in a CE phase, the CE masses typically
lie in the range of a few $10^{\rm -4}~M_{\odot}$ to a few
$10^{\rm -3}~M_{\odot}$. It is difficult to detect this
hydrogen-rich material directly. However, such low-mass CEs might
leave footprints in the high-velocity features in the spectrum of
SNe Ia (\citealt{MAZZALI05a,MAZZALI05b}). In a few cases, the CE
mass can be larger than $0.1~M_{\odot}$, even as large as
$1~M_{\odot}$.  Such explosions may show the properties of
2002ic-like SNe.  The results here are similar to those in
\citet{MENGXC17}, except that the CE for some systems may be as
massive as $2~M_{\odot}$.  We show the results of our evolutionary
calculations for $M_{\rm WD}^{\rm i}=1.3~M_{\odot}$ as an example
in Fig.~\ref{grid}, where most of the systems will explode in the
CE phase. Especially for the systems in the upper-right region,
the CE mass is generally larger than $0.1~M_{\odot}$ when $M_{\rm
WD}=1.378~M_{\odot}$.

In Fig.~\ref{cour130}, we show the contours leading to SNe Ia for
different initial WD masses. In the figure, the triple-dot-dashed
line approximately marks the boundary between $M_{\rm
CE}\geq0.1~M_{\odot}$ and $M_{\rm CE}<0.1~M_{\odot}$ when $M_{\rm
WD}=1.378~M_{\odot}$. This clearly shows that a massive WD with
$M_{\rm WD}^{\rm i}\geq1.1~M_{\odot}$ is required for the system
to explode with a massive CE. Moreover, only when the initial WDs
are massive enough, i.e.\ $M_{\rm WD}^{\rm i}=1.3~M_{\odot}$, may
the explosions show hybrid properties of both core-collapse SNe
and SNe Ia.

\begin{figure}
\centerline{\includegraphics[angle=270,scale=.35]{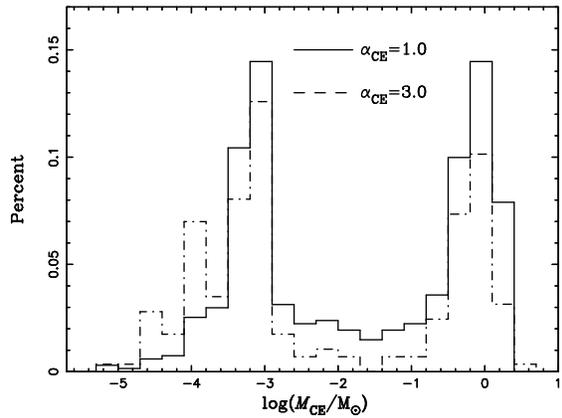}}
\caption{The distribution of CE masses when $M_{\rm
WD}=1.378~M_{\odot}$ for different values of $\alpha_{\rm
CE}$.}\label{mcedis}
\end{figure}

\subsection{The distribution of $M_{\rm CE}$}\label{sect:3.3}
Since most hybrid CONe WDs will explode in the CE phase, we
performed two BPS simulations with different values of the CE
ejection parameter $\alpha_{\rm CE}$ to obtain the CE mass
distribution when $M_{\rm WD}=1.378~M_{\odot}$
(Fig.~\ref{mcedis}). In Fig.~\ref{mcedis}, the distribution of CE
masses shows two peaks for either case, i.e. one around
$\sim1~M_{\odot}$ and one around $\sim10^{\rm -3}~M_{\odot}$. The
peaks can be explained by the different evolutionary stages and
differences in mass-transfer rates at the onset of RLOF (see the
detailed explanation in \citealt{MENGXC17}). It may be difficult
to detect a CE of $\sim10^{\rm -3}~M_{\odot}$ directly, but the CE
as massive as $\sim1~M_{\odot}$ will be directly detectable as
shown in the spectra of some SNe Ia-CSM (\citealt{SILVERMAN13}).
Among all the SNe Ia with hybrid CONe WDs, about 25\%--40\% SNe Ia
explode in massive CEs (note that Fig.~\ref{mcedis} only includes
systems exploding in a CE phase, not those in SSS or RN phases).
Because of the high initial WD masses, SNe Ia from hybrid CONe WDs
originate from a relatively young population, i.e.\ with ages less
than 1 Gyr (see also \citealt{MENGXC14}).

As discussed above, hybrid CONe WDs may explode in different
environments, i.e.\ in a CE phase with envelopes of different
masses, an SSS phase, or a RN phase, leading to differences in the
supernova properties. Considering that both SN 2002cx-like and SN
Ia-CSM events show SN 1991T-like spectra and are hosted in
late-type galaxies and that numerical simulations of the
explosions of hybrid CONe WDs can reproduce the properties of
2002cx-like SNe (\citealt{LIWD02}; \citealt{SILVERMAN13};
\citealt{KROMER15}), we propose that both subclasses originate
from the explosion of hybrid CONe WDs in SD systems, where those
exploding with massive CEs are associated with SN Ia-CSM events,
while those exploding in less massive CEs or in a SSS/RN phase
correspond to SN 2002cx-like SNe. In other words, SNe Ia-CSM have
a denser CSM than 2002cx-like SNe. In addition, since most SNe Ia
involving CO WDs explode in SSS/RN phases (\citealt{MENGXC17}),
our CEW model predicts that both SN Ia-CSM and SN 2002cx-like
events statistically have a denser CSM than normal SNe Ia. At
present, no such CSM density sequence has been reported
observationally, but based on the analysis of a SN Ia sample
without radio detections, an upper-limit sequence on the CSM
density indeed exists around SN Ia-CSM, 2002cx-like SNe and normal
SNe Ia (\citealt{CHOMIUK16}). Future X-ray observations could
provide more meaningful constraints on such a density sequence
since X-ray observations are more sensitive than radio
observations for CSM detections (\citealt{MARGUTTI12,MARGUTTI14};
\citealt{MENGXC16}).

Specifically, some SN Ia-CSM events could originate from the
explosions of single stars as suggested in
Section.~\ref{sect:3.1}, and show the hybrid properties of
core-collapse SNe and SNe Ia as seen, e.g., in SN 2012ca
(\citealt{INSERRA14,INSERRA16}); however, such objects must be
very rare for higher initial WD and companion masses according to
the initial mass function of stars (\citealt{MS79}). In fact, in
our BPS simulations with $10^{\rm7}$ binaries, no such object was
produced: this would imply that such objects contribute less than
$\sim0.03\%$ to all SNe Ia. All binary systems in our study must
have experienced a dynamical unstable CE and spiral-in phase in
the past to form a WD + MS system. After the CE phase, binary
systems with relatively massive WDs and companions tend to have
relative long orbital periods and produce SNe Ia exploding in
massive CEs, i.e.\ are SN Ia-CSM objects, rather than hybrid
supernovae.

Here, we do not show the evolution of the SN Ia birth rate with time,
since the evolution is very similar to that shown in
\citet{MENGXC14}. However, because the parameter space leading to SNe
Ia from the CEW model is larger than that from the OTW model, the
birth rate from the CEW model is generally higher than that from the
OTW model by $\sim30\%$ (\citealt{MENGXC17}). Similarly, the birth
rate here is also higher than that in \citet{MENGXC14} by roughly
$10\%$ to $27\%$, depending on the adopted value for $\alpha_{\rm
  CE}$. Considering the various uncertainties discussed in
\citet{MENGXC14}, we obtain a conservative upper limit for the
contribution to all SNe Ia from hybrid CONe WDs of the order of
$~10\%$ (as estimated by \citealt{MENGXC14}). In addition,
comparing this study with the results for CO WD + MS systems in
\citet{MENGXC17}, we can estimate the number ratio of peculiar SNe
Ia from hybrid CONe WDs to normal ones from CO WDs to be between
1/5 and 1/9, depending on the $\alpha_{\rm CE}$ value.  As
expected, this ratio is probably larger than compatible with
observations, e.g.\ 1/14 in \citet{LIWD11} and 1/5-1/47 in
\citet{GRAUR16}, since only CO WD + MS systems are considered to
produce normal SNe Ia here.

\section{DISCUSSIONS AND CONCLUSIONS}\label{sect:4}
\subsection{The contribution of SNe Ia from hybrid CONe WDs to all SNe Ia}\label{sect:4.1}
In this paper, we have pointed out that SN 2002cx-like and SN Ia-CSM
objects share two common properties, i.e.\ both subclasses of
supernovae show a SN 1991T-like early spectrum and are hosted by
late-type galaxies, and argued that they could share the same
progenitor origin. Considering that the simulated explosions of hybrid
CONe WDs appear to explain the properties of 2002cx-like SNe (see the
discussions in \citealt{MENGXC14}; \citealt{KROMER15};
\citealt{BRAVO16}), we propose that both SN 2002cx-like and SN Ia-CSM
objects originate from the explosions of hybrid CONe WDs in SD
systems, based on the CEW model of \citet{MENGXC17}. The BPS
calculations show that, depending on the value of $\alpha_{\rm CE}$,
roughly 25\%-40\% of SNe Ia with hybrid CONe WDs will show the
properties of SN Ia-CSM objects. Hence, the predicted number ratio of
SNe Ia-CSM to SN 2002cx-like objects is in the range of $\frac{1}{3}$
to $\frac{2}{3}$.

\citet{FOLEY13} and \citet{SILVERMAN13} summarized the samples of
SN 2002cx-like and SN Ia-CSM objects before 2013, respectively. In
their samples, there are 25 SN 2002cx-like and 16 SN Ia-CSM
objects. Hence, the observed number ratio of SNe Ia-CSM to SN
2002cx-like objects is $0.64^{\rm +0.23}_{\rm -0.18}$, where the
error bars assumed a binomial distribution (\citealt{CAMERON11}).
Since some SNe IIn could be misclassified as SNe Ia-CSM
(\citealt{BENETTI06}; \citealt{INSERRA14}) and SN Ia-CSM objects
are generally brighter than 2002cx-like SNe, making them more easy
to discover, the observed number ratio is likely an overestimate.
Therefore, our predicted and the observed number ratios appear
consistent with each other, at least within the observational
errors. On the other hand, since the explosions of hybrid CONe WDs
display significant variability, e.g., the range of $^{\rm 56}$Ni
yields from hybrid CONe WDs is much wider than that from CO WDs
(\citealt{KROMER15}; \citealt{WILLCOX16}), the properties of SN
2002cx-like and SN Ia-CSM objects would also appear more
heterogeneous than normal SNe Ia, consistent with the diversity of
observed 2002cx-like SNe (\citealt{FOLEY13}).

The contribution of SNe Ia with hybrid CONe WD of all types is
likely less than $\sim10\%$ of the total SN Ia rate. Although
estimates for the fraction of 2002cx-like SNe have varied
significantly from $\sim5\%$ to $\sim30\%$ (\citealt{LIWD11};
\citealt{FOLEY13}), recent analyses based on volume-limited
samples appear to favour a lower value around 5\,\% (see, e.g.,
\citealt{WHITE15}; \citealt{ASHALL16}; \citealt{GRAUR16}). One
possibility for the large uncertainty of the contribution of
2002cx-like SNe is that the subclass of 2002cx-like SNe has a
different origin since the subclass presents quite significant
heterogeneity, e.g., their peak absolute magnitude ranges from
$M_{\rm V}\sim -14$ to $M_{\rm V}\sim -19$, which is much larger
than for other SNe Ia. Conceivably, the faint subclass to which SN
2008ha belongs originates from helium deflagrations or detonations
on CO WDs rather than hybrid CONe WDs (\citealt{WANGB13};
\citealt{NEUNTEUFEL17}). If that were the case, our model would
only contribute to part of the 2002cx-like SN subclass.

The contribution of SN Ia-CSM objects to all SNe Ia is still
unclear. SN 2002ic-like SNe contribute about 1\% to all SNe Ia
(\citealt{ALDERING06}), but the contribution of SN Ia-CSM objects
to all SNe Ia is higher than that of SN 2002ic-like SNe since SN
1997cy and SN 1999E have also been classified as SN Ia-CSM
objects. Considering that the number of discovered SN Ia-CSM
objects is smaller than that of 2002cx-like SNe and that SN Ia-CSM
objects are relatively more easily discovered because of their
higher luminosities, the contribution of SN Ia-CSM objects to all
SNe Ia is probably less than that of 2002cx-like SNe. We conclude
that SN Ia-CSM objects roughly contribute between 1\% and 5\% of
all SNe Ia. This estimate is also consistent with our estimate on
the ratio of SNe Ia with and without massive CEs. Therefore, the
contribution of 2002cx-like and SN Ia-CSM objects to all SNe Ia is
also consistent with our estimate for SNe Ia with hybrid CONe WDs.


\subsection{The scale of the CSM}\label{sect:4.2}
Fig.~\ref{mcedis} shows the distribution of CE masses at the
moment when $M_{\rm WD}=1.378~M_{\odot}$. As we argued before, the
interaction of the supernova ejecta with the CSM produced by
massive CEs may partly contribute to the high luminosity of SNe
Ia-CSM. Observations suggest that there can be a time delay
between the supernova explosion and the interaction with the CSM
of up to tens of days after the explosion (\citealt{HAMUY03};
\citealt{ALDERING06}; \citealt{DILDAY12}; \citealt{SILVERMAN13};
\citealt{SOKER17}). This seems to require a time delay between the
moment when $M_{\rm WD}=1.378~M_{\odot}$ and the supernova
explosion, and that the time delay must be long enough to eject
the CE in some cases. Using our estimates for the CE mass and the
mass-loss rate when $M_{\rm WD}=1.378~M_{\odot}$, we find that it
will take $10^{\rm 4}$\,yr to a few $10^{\rm 5}$\,yr to eject the
CE. A possible mechanism for the time delay is the
spin-up/spin-down model (\citealt{JUSTHAM11};
\citealt{DISTEFANO12}), in which a rapidly rotating
super-Chandrasekhar WD experiences a spin-down phase before the
supernova explosion.  Although the spin-down time scale, is
currently not yet understood from a purely theoretical point of
view, empirically it has been estimated to be between $10^{\rm 5}$
yr and a few $10^{\rm 7}$ yr, much longer than the timescale
required to eject the CE (\citealt{DISTEFANO12};
\citealt{MENGXC13}).

The spatial scale of the CSM formed by the final dissipation of a
CE depends on the spin-down timescale,$\tau_{\rm sd}$, and the
wind velocity, $v_{\rm w}$.  The wind velocity in the CEW model is
uncertain but is plausibly in the range of 10 $-$ 100 km/s, likely
higher than the typical wind velocity of AGB stars due to the
extra gravity from the companion, e.g. a higher escape velocity
from the surface of the CE (\citealt{MENGXC17}). A spin-down
timescale of a few $10^{\rm 5}$ yr and a wind velocity of 10 km/s
would put the outer boundary of the CSM at a distance of $v_{\rm
w}\times\tau_{\rm sd}\ga 10^{\rm 5}$ AU. The inner boundary
depends on how the CE is dissipated. As the CE is much larger than
the final orbit, some of it is likely to remain bound and
ultimately form a circumbinary disc-like structure (perhaps
similar to what is seen around AGB binaries; e.g.\
\citealt{BUJARRABAL13}).
The inner boundary could then be anywhere between a few times the
final orbital separation and several AU, i.e.\ the radius of the
CE (depending on the angular momentum stored in the remnant CE).
Simply taking a terminal ejecta velocity of 30000 km/s
(\citealt{WANGLF06}), the ejecta may interact with the CSM
immediately or within one day. Similarly, assuming a spin-down
timescale of a few $10^{\rm 6}$ yr and a wind velocity of 100
km/s, the ejecta may catch up with the CSM within tens of days
after the explosion. The estimate of the onset time of the
interaction is consistent with observations
(\citealt{SILVERMAN13}). Therefore, to explain the time delay
between the supernova explosion and the interaction with the CSM,
a spin-down timescale of $\sim10^{\rm 6}$ yr is favoured which is
consistent with the constraint derived in \citet{MENGXC13}.

The estimate for the spatial scale of the CSM for SN Ia-CSM
objects is based on the CE mass and mass-loss rate when $M_{\rm
WD}=1.378~M_{\odot}$. This mass-loss rate is limited by the
Eddington luminosity of WDs in the standard CEW model. However,
during the spin-down phase, it is possible in principle to drive a
super-Eddington wind and hence a higher mass-loss rate, leading to
a denser CSM. On the other hand, the companion may continue to
feed material to the CE during the spin-down phase, and therefore
the final amount of the CSM around SN Ia-CSM objects could be
larger than that shown in Fig.~\ref{mcedis}.

While the mass of the CSM around SN Ia-CSM objects from
observations is model-dependent, it may cover a large range since
the total luminosity and the decline rate of light curves differ
significantly for different SNe Ia-CSM: see, e.g., SN 2005gj,
2002ic and PTF 11kx (\citealt{ALDERING06}; \citealt{HAMUY03};
\citealt{DILDAY12}). In our CEW model, the amount of CE mass
ranges from $\sim0.1~M_{\odot}$ to more than $2~M_{\odot}$,
peaking around $1~M_{\odot}$. The distribution provides a
potential statistical test for our SN Ia-CSM model.

The detailed shape of the CSM in the CEW model is highly uncertain
as it depends on numerous uncertainties in the current version of
the model.  A structure similar to what is seen in planetary
nebulae (PNe) is possible, maybe with multiple thin shells as are
often seen in PNe due to binary effects (\citealt{MASTRODEMOS99};
\citealt{SOKER05}). In addition, for a Chandrasekhar-mass WD,
RN-like eruptions may occur and lead to the formation of
multiple-shell structures in the CSM. Observationally, some SNe Ia
are associated with planetary nebulae, e.g, exploding inside PNe
(\citealt{TSEBRENKO15}; \citealt{CIKOTA17}), and some supernova
remnants even show a special `ear' structure that may be expected
from CE evolution (\citealt{TSEBRENKO15}). Our model could explain
these observations in principle and especially the multiple thin
shells detected in PTF11kx, which had previously been associated
with a symbiotic system (\citealt{DILDAY12}).

\subsection{SNe Ia-CSM}\label{sect:4.3}
Besides belonging to a young population and showing a 1991T-like
spectrum, SNe Ia-CSM have relatively long light curve rise times,
large peak luminosities, and potentially show evidence for dust
formation at late times (\citealt{SILVERMAN13}). Our model may
potentially explain all these characteristics of SNe Ia-CSM. In
our model, SNe Ia-CSM are the explosions of hybrid CONe WDs with
massive CEs, in which strong hydrogen emission lines are predicted
from the collision of the supernova ejecta with the CSM formed
from the CE. At the same time, with the expansion and cooling of
the CE, dust may form around the progenitor system
(\citealt{LUGL13}). Such dust formation in the circumstellar
environment is also observed in some normal SNe Ia
(\citealt{WANGLF05}; \citealt{GOOBAR08};
  \citealt{JOHANSSON13}).

For SNe Ia-CSM, the high luminosity is very likely correlated with
the long rise time as has been demonstrated for 2002cx-like SNe Ia
(\citealt{MAGEE16}). However, for SNe Ia-CSM, the origin of the
correlation could be more complex.  Definitely, the collision
between supernova ejecta and the CSM partly contribute to the high
luminosity and long rise time of SNe Ia-CSM (\citealt{HAMUY03}).
In addition, the high peak luminosity could imply a high
production of $^{\rm 56}$Ni in SNe Ia-CSM. For example, if SNe
Ia-CSM originate from the hybrid CONe WDs with massive CEs, i.e.
the systems with massive companions, the CONe WDs experience a
relatively shorter cooling time from the birth of the WDs to the
onset of accretion than those without massive CE due to their more
massive companion (cf.\ Fig.~\ref{cour130}), i.e.\ the average
cooling times for SNe with and without massive CEs are 0.21 Gyr
and 0.44 Gyr, respectively. Such a difference of average cooling
times could even imply that SNe Ia-CSM tend to trigger central
ignition, while 2002cx-like SNe Ia may favor off-center ignition
(\citealt{CHENXF14}). Moreover, the shorter cooling time leads to
a lower central density when the explosion occurs
(\citealt{PODSIADLOWSKI08}; \citealt{CHENXF14}). A lower central
density might mean a smaller region suitable to trigger explosive
carbon ignition in the center of the CONe WDs, which results in a
larger production of $^{\rm 56}$Ni, even higher than from CO WDs
(\citealt{WILLCOX16}).

\subsection{2002cx-like supernovae}\label{sect:4.4}
Recently, \citet{JHA17} summarized the properties of 2002cx-like
SNe: 1) the population forms quite a heterogeneous class, e.g.\
spanning a wide range of peak photometric luminosities; 2)
2002cx-like SNe have a spectrum similar to SN 1991T, but a
luminosity similar to SN 1991bg; 3) they are preferentially found
in late-type galaxies, indicating a young population; 4) they are
characterized by a lower ejecta velocity compared with normal SNe
Ia, and 5) 2002cx-like SNe Ia are likely to contribute to about
5\% of all SNe Ia (although this estimate is very uncertain).

SNe Ia from hybrid CONe WDs could in principle display significant
heterogeneity, similar to what is seen in 2002cx-like SNe Ia. For
one thing, the difference of the CO core mass in the hybrid CONe
WDs could partly contribute to the difference in their peak
luminosity, since the carbon abundance may affect the brightness
of SNe Ia (\citealt{NOM03}). In addition, whether a detonation
develops after the initial deflagration phase or not may also
contribute significantly to their heterogeneity. Even if a
detonation develops, the explosions of hybrid CONe WDs also show
differences depending on the different CO core mass and the
different number of ignition regions (\citealt{BRAVO16};
\citealt{WILLCOX16}). For a lower carbon abundance compared with
normal CO WDs, SNe Ia from hybrid CONe WDs may have a lower
luminosity and lower ejecta velocities (\citealt{NOM03};
\citealt{BRAVO16}), consistent with 2002-cx-like SNe Ia. In
addition, SNe Ia from hybrid CONe WDs are younger than 1 Gyr --
they may even be as young as 30 Myr, consistent with 2002cx-like
SNe Ia (\citealt{MENGXC14}). Our estimated rate of SNe Ia from
hybrid CONe WDs match the rate of 2002cx-like SNe (see section
\ref{sect:4.1}).  Therefore, the properties of 2002cx-like SNe Ia
could -- in principle -- be reproduced by our model.

Observationally, some 2002cx-like SNe Ia show evidence for helium
in their spectra (\citealt{FOLEY13}). Based on their detailed BPS
simulations, \citet{MENGXC14} found that some MS companions become
helium-rich when the CONe WD + MS systems form, which could
explain the observed helium in some 2002cx-like SNe Ia.  In
addition, in the case of SN 2012Z, pre-supernova observations
showed the presence of a luminous blue object that did not
disappear after the explosion (\citealt{MCCULLY14};
\citealt{JHA17}); this is consistent with the CONe WD + He star
channel (\citealt{WANGB14}), although BPS simulations do not
reproduce the system properties in detail.  In our CEW model, the
companion may become almost a `naked' helium core after envelope
ejection, and the helium core may be as massive as
$0.97~M_{\odot}$ (see Fig. 4 and discussions in
\citealt{MENGXC17}). If helium is ignited in the companion's
center before the supernova explosion, such a companion could fit
the observations of the star associated with SN 2012Z.  Whether
the companion may reproduce the properties of the star associated
with SN 2012Z or not will heavily depend on the spin-down
timescale, an issue we will address in detail in the future.

\subsection{Hybrid/single supernovae}\label{sect:4.5}
In this paper, we suggest a new kind of thermonuclear explosion
from single CONe WDs in an AGB-like CE, i.e.\ a SN I 1/2 in the
nomenclature of \citet{IBEN83}, which shows hybrid properties of
CC SNe and SNe Ia. In the case where all the envelope is ejected
before the supernova explosion, the explosion could be from a {\em
single} CONe WD. Although the detailed properties of such
explosions are difficult to ascertain, we can still speculate on
some of their unusual charactersitics. First, compared with the
normal AGB stars, the mergers would experience a higher mass-loss
rate as it is enhanced by rapid rotation (\citealt{SOKER98};
\citealt{POLITANO08}); this is helpful for explaining why the
progenitors of some SNe Ia-CSM experience very high mass loss
(\citealt{SILVERMAN13}). The amount of CSM mass around such SNe
may be as high as $\sim 3 M_{\odot}$, and the CSM could possibly
take the form of a dense torus or disk structure
(\citealt{TAAM10}; \citealt{OHLMANN16}), as deduced from X-ray
observations in the case of SN 2012ca (\citealt{BOCHENEK17}). In
addition, since the CONe WDs may continue to increase their mass
during the spin-down phase, the exploding CONe WD could be
super-Chandrasekhar objects. Interestingly, the late-time optical
and infrared spectra of SN 2012ca show low [Fe III]/[Fe II] ratios
and strong [Ca II] lines, which are characteristics of the
super-Chandrasekhar-mass candidate SN 2009dc
(\citealt{TAUBENBERGER13}; \citealt{FOX15}). Finally, for hybrid
explosions originating from the merger of a CONe WD + MS system,
no surviving companion would exist in the supernova remnant after
the supernova explosion, which in principle could explain the fact
that no surviving companion has yet convincingly been identified
in any supernova remnant (\citealt{SCHAEFER12};
\citealt{KERZENDORF12,KERZENDORF14}). However, such special SNe Ia
are rare and only contribute to less than 0.03\% of all SNe Ia
according to our current BPS simulations and are therefore
unlikely to explain the lack of identified
surviving companions in the majority of cases.  \\

In summary, we propose that both SN 2002cx-like and SN Ia-CSM
objects share the same origin, i.e. from the hybrid CONe WD + MS
systems. In addition, we predict a hybrid object which shows
features of both SNe II and SNe Ia, but without a surviving
companion. As a final caveat we note that this discussion is based
on the CEW model as presented in \citet{MENGXC17}, which is still
under development (see the detailed discussions in
\citealt{MENGXC17}), and the model still needs to be verified by
observations. On the other hand, if future observations support
the suggestion proposed in this paper, it could provide support
for the CEW model.

\section*{Acknowledgments}
This work was partly supported by Natural Science Foundation of
China (Grant Nos. 11473063, 11522327 and 11733008), Yunnan
Foundation (No. 2015HB096, 11733008), CAS (No. KJZD-EW-M06-01),
CAS ``Light of West China'' Program and Key Laboratory for the
Structure and Evolution of Celestial Objects, Chinese Academy of
Sciences.

\end{document}